\documentclass[letterpaper,twocolumn,10pt]{article}
\usepackage{usenix,epsfig,endnotes}
\pdfoutput=1
\usepackage{times}
\usepackage{xspace}
\usepackage{algcompatible}
\usepackage{algorithm}
\usepackage[noend]{algpseudocode}

\usepackage{amssymb}
\usepackage[colorlinks = true,
                     citecolor = blue,
                     urlcolor = blue]{hyperref}
\usepackage{comment}
\usepackage{multirow}
\usepackage{tabulary}
\usepackage{paralist}
\usepackage{enumitem} 
\usepackage{url}
\usepackage{relsize}
\usepackage{graphicx}
\usepackage{flushend}
\usepackage{setspace}
\usepackage[font=small]{caption}

\setlist[itemize]{itemsep=-0.3em, topsep=2pt, leftmargin=\parindent}
\setlist[enumerate]{itemsep=-0.3em, topsep=2pt, leftmargin=\parindent}

\newcommand{\thesystem}{Eunomia\xspace}
\newcommand{\theprototype}{EunomiaKV\xspace}
\newcommand{\thefiguresize}{0.45\textwidth}
\newcommand{\themotivfiguresize}{0.48\textwidth}

\newtheorem{property}{Property}

\begin{document}


\title{\Large \bf Unobtrusive Deferred Update Stabilization for Efficient Geo-Replication}
\author{Chathuri Gunawardhana$^1$, Manuel Bravo$^{1,2}$ and Lu\'is Rodrigues$^1$ \\
\small {\em  $^1$INESC-ID, Instituto Superior T\'ecnico, Universidade de Lisboa \quad
          $^2$Universit\'e Catholique de Louvain, Belgium} \\ [2mm]
}
\date{}
\maketitle
\begin{abstract}

  In this paper we propose a novel approach to manage the throughput vs
  latency tradeoff that emerges when managing updates in
  geo-replicated systems. Our approach consists in allowing full
  concurrency when processing local updates and using a deferred
  \emph{local} serialisation procedure before shipping updates to
  remote datacenters. This strategy allows to implement inexpensive
  mechanisms to ensure system consistency requirements while avoiding
  intrusive effects on update operations, a major performance
  limitation of previous systems. We have implemented our approach as
  a variant of Riak KV. Our extensive evaluation shows that we
  outperform sequencer-based approaches by almost an order of
  magnitude in the maximum achievable throughput. Furthermore, unlike
  previous sequencer-free solutions, our approach reaches nearly
  optimal remote update visibility latencies without limiting
  throughput.

\end{abstract}

\section{Introduction}

Geo-replication is a requirement for modern internet-based
services in order to improve user-perceived latency. Unfortunately,
due to the long network delays among
sites, synchronous replication is prohibitively slow for most
practical purposes. Therefore, many systems resort to weaker
consistency semantics that permit some form of asynchronous replication
strategy.


Among the many consistency
guarantees that allow for asynchronous replication\,\cite{hat}, causal
consistency\,\cite{causalmemory} has been
identified as the strongest consistency model that an always-available
system can implement\,\cite{limitations, cac}, becoming of practical
relevance in geo-replicated settings. In fact, causal consistency is
key in many geo-replicated storage systems offering from
weak\,\cite{bayoulog, cops, chainreaction, swiftcloud} to strong
consistency guarantees\,\cite{psi, redblue, indigo}.


Unfortunately, implementing causal consistency is costly due to the
computation, communication, and storage overhead caused by metadata
management\,\cite{ourpaper, rachid, potential}. A common solution to reduce this
cost consists in compressing metadata by serializing sources of
concurrency, which unavoidably creates \emph{false
  dependencies} among concurrent events, increasing
\emph{visibility latencies} (time
interval between the instant in which an update is installed in
its origin datacenter and when it becomes visible in remote
datacenters).

To safely compress metadata, designers of causally consistent systems
rely either on: (i) centralized \emph{sequencers} (commonly one per
datacenter)\,\cite{swiftcloud, chainreaction}; or (ii) \emph{global
  stabilization} procedures\,\cite{gentlerain, cure} (executed across
datacenters). The former has the advantage of making trivial---and
therefore inexpensive---the dependency checking procedures at the cost
of severely limiting concurrency, as sequencers operate in the
critical path of clients. On the contrary, the latter avoids
centralized synchronization points at the cost of periodically running
a global stabilization procedure in the background. The cost of this
procedure has pushed some systems to over-compress metadata to avoid
impairing throughput, with a significant penalty on the visibility
latencies\,\cite{gentlerain}.

In this paper, we propose, implement, and evaluate a novel approach
to address the metadata size versus visibility latency tradeoff. Our
approach  has some similarities with systems that rely
on global stabilization but also significant differences. As
with\,\cite{gentlerain, cure}, we let local updates proceed without
any a priori synchronization. However, unlike previous systems, we
totally order all updates, in a manner consistent with causality,
before shipping updates to remote datacenters. As a result, 
expensive global stabilisation is avoided, as it is trivial for a datacenter to
check whether all updates subsumed in the timestamps piggybacked by
remote updates have been locally applied (similarly to
sequencer-based solutions).

We have implemented our approach as a variant of the open source
version of Riak KV\,\cite{riakkv}. We have augmented Riak with a service that totally
orders all the updates, before shipping them, that we have called
\thesystem\footnote{Greek goddess of law and legislation, her name can
  be translated as "good order".}. Our experimental results
show that Riak+\thesystem outperforms sequencer-based systems by almost an order
of magnitude while serving significantly better quality-of-service to
clients compared with systems based on global stabilisation
  procedures.

  In summary, the contributions of this paper are the following: i)
  The introduction of \thesystem, a new service for unobtrusively
  ordering updates (\S\ref{sec:thesystem}); ii) A fault tolerant
  version of \thesystem~(\S\ref{sec:faulttolerance}); iii) A sound
  experimental comparison of the maximum load that traditional
  sequencers and the newly introduced \thesystem can handle, and their
  potential bottlenecks (\S\ref{sec:upperbound}); iv) The Integration
  of \thesystem into an always-available geo-replicated data store
  (\S\ref{sec:georeplication}) and its performance comparison to
  state-of-the-art solutions (\S\ref{sec:expstore}).


\section{Motivation and Goals}

We start by motivating our work with a
simple experiment, showing that: (i) the major throughput impairment of
sequencer-based solutions is the fact that they operate in the
critical path of clients; and (ii) global stabilization procedures are
expensive in practice, forcing designers to favour either throughput
or visibility latencies.

\begin{figure}[t]
\centering
    \includegraphics[width=\themotivfiguresize]{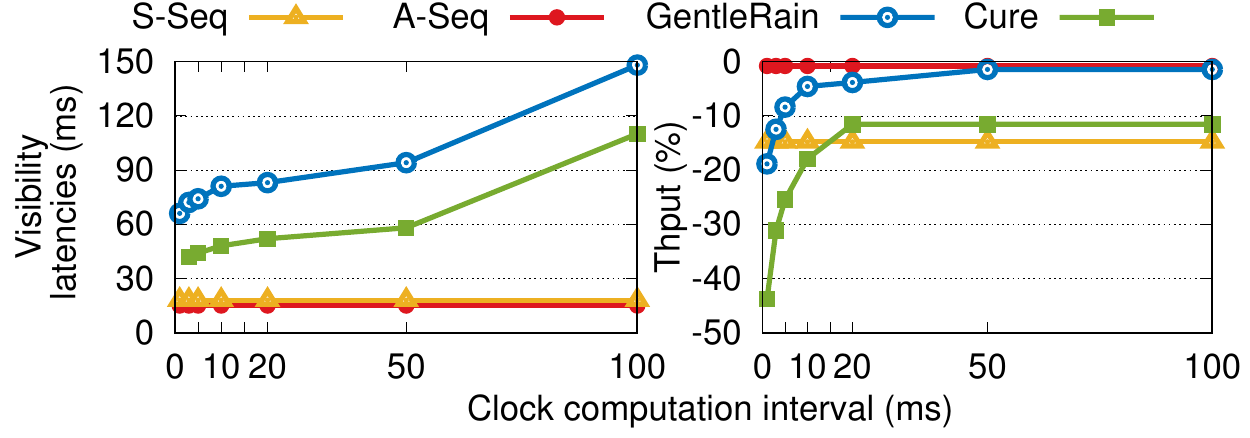}
    \caption{Update visibility latency vs throughput  tradeoff.}
    \label{fig:motivation}
\end{figure}

Figure~\ref{fig:motivation} plots the throughput penalty and
visibility latency overhead introduced by state-of-the-art causally
consistent solutions. Results are normalized against an eventually
consistent system, which adds no overhead due to consistency
management. We vary from $1ms$ to $100ms$ the interval between global
stabilization computations to better understand the cost and the
consequences of such mechanism. Our deployment consists of 3
datacenters. The round-trip-times across datacenters are $80ms$
between datacenter 1 ($dc_1$) and both $dc_2$ and $dc_3$; and $160ms$
between $dc_2$ and $dc_3$. In the figure (left plot), latencies refer
to the ($90^{th}$ percentile) delays incurred by each system at $dc_2$
for updates originating at $dc_1$. We compare the performance of 4
systems, namely \emph{S-Seq}, \emph{A-Seq}, \emph{GentleRain} and
\emph{Cure}.

\emph{S-Seq} is a system that relies on a sequencer per datacenter to
compress metadata; it uses a vector with an entry per datacenter to track causality, as in\,\cite{chainreaction, swiftcloud}.
\emph{A-Seq} is an asynchronous (bogus) variant of \emph{S-Seq}, that
contacts the sequencer in parallel with applying the
update. \emph{A-Seq} does the same total amount of work as
\emph{S-Seq} and, although it fails to capture causality, it serves 
to reason about the potential benefits of removing sequencers from
client's critical operational
path. \emph{GentleRain}\,\cite{gentlerain} and
\emph{Cure}\,\cite{cure} are well known solutions that rely on global
stabilization. The former favours throughput, over-compressing
metadata into a single scalar; the latter favours visibility
latencies, compressing metadata into a vector with an entry per
datacenter.

The results confirm that the costs inherent to global stabilization
force designers to choose between optimizing throughput and visibility
latencies. As Figure~\ref{fig:motivation} shows, Cure offers lower
visibility latencies than GentleRain (as causality is more precisely
tracked) at the cost of penalizing throughput. GentleRain does the
opposite tradeoff favouring throughput. Cure can tune this tradeoff by
choosing longer intervals among global stabilization
occurrences. Nevertheless, even with long intervals ($100ms$), Cure
still significantly degrades system throughput by
11.6\%. Interestingly, results also show that by removing the
sequencer from client's critical operational path, sequencer-based
approaches could potentially pick a better spot in the tradeoff space,
by providing throughput and visibility latencies comparable to
GentleRain and Cure respectively, with almost no performance overhead
when compared to the baseline. Note that in the above experiment,
sequencers are not overloaded; therefore, the throughout penalty
(14.8\%) is exclusively caused by the synchronous communication with
the sequencer at every client update operation. Later, in
\S\ref{sec:upperbound}, we experimentally measure the maximum load
that sequencers can handle.

From these results, it is possible to get the following insight: in
order to alleviate the tension between throughput and visibility
latencies, one has to (i) avoid global stabilization, and (ii) rely on
an abstraction similar to sequencers that allows for
trivial---therefore inexpensive---dependency checking procedures,
while removing its operation from the client's critical path. Our goal
was then to design \thesystem,  a system with such characteristics.

\section{\thesystem: Unobtrusive Ordering}
\label{sec:thesystem}

In this section, we present the design and rationale underlying
\thesystem, a new service conceived to replace sequencers as building
blocks in weakly consistent geo-replicated storage
systems. 
Unlike traditional sequencers,
\thesystem lets local client operations to execute without synchronous
coordination, an essential characteristic to avoid limiting
concurrency and increasing operation latencies. Then, in the
background, \thesystem establishes a serialization of all updates
occurring in the local datacenter in an order consistent with
causality, based on timestamps generated locally by the individual
servers that compose the datacenter. We refer to this process as
\emph{site stabilization procedure}. Thus, \thesystem is capable of
abstracting the internal complexity of a multi-server datacenter
without limiting the concurrency. Therefore, \thesystem can be used to
improve any existing sequencer-based solution to enforce causal
consistency across geo-locations \,\cite{bayoulog, swiftcloud,
  chainreaction}, as shown in \S\ref{sec:georeplication}.

\subsection{\thesystem Into Play}
\label{sec:protocol}

In order to convey  how \thesystem works, we start by presenting the
 protocol used to support the interaction between \thesystem and the machines that
constitute a datacenter. In the  exposition, we assume that the
key-space is divided into $N$ \textit{partitions} distributed among datacenter machines.
Updates to items belonging to the same partition are serialized by the native update
protocol. 
We assume FIFO links among partitions and
\thesystem. Table~\ref{table:definitions} provides a summary of the
notation used in the protocols.

\begin{table}[t!]
\centering
\footnotesize
\renewcommand{\arraystretch}{1}
  \begin{tabulary}{\columnwidth}{@{} R | L @{}}
    $N$ & Number of partitions \\
\hline
    \textit{Clock}$_c$ & Client $c$ clock \\
\hline
    $p_n$ & Partition $n$ \\
    \textit{Clock}$_n$ & Current physical time at $p_n$ \\
\hline
    \textit{Ops} & Set of unstable operations at \thesystem\\
    \textit{PartitionTime }& Vector with an entry per partition at
              \thesystem\\
\hline
    $u_j.ts$ & Timestamp assigned to update $u_j$\\
  \end{tabulary}
\caption{Notation used in the protocol description.}
\label{table:definitions}
\end{table}

\begin{algorithm}[t]
  \begin{algorithmic}[1]
\footnotesize
	\Function{read}{\textit{Key}}
        \State {\bf send} \Call{read}{\textit{Key}} {\bf to}
        \Call{responsible}{\textit{Key}}\label{clientread}
        \State {\bf receive} $\langle$\textit{Value}, \textit{Ts}$\rangle$ {\bf from} \Call{responsible}{\textit{Key}}
        \State \textit{Clock}$_c\gets$ \Call{max}{\textit{Clock}$_c$, \textit{Ts}}\label{maxclient} 
        \State \Return \textit{Value}
    \EndFunction
    \vspace{1mm}
\Function{update}{\textit{Key}, \textit{Value}}
        \State {\bf send} \Call{update}{\textit{Key},
          \textit{Value}, \textit{Clock}$_c$} {\bf to} \Call{responsible}{\textit{Key}}\label{clientupdate}
        \State {\bf receive} \textit{Ts} {\bf from} \Call{responsible}{\textit{Key}}
        \State \textit{Clock}$_c \gets$ \textit{Ts}\label{tsupdate}
        \State \Return \textit{ok}
    \EndFunction
  \end{algorithmic}
  \caption{Operations at client $c$}
  \label{alg:client}
\end{algorithm}

\thesystem assumes that each individual partition can assign a
timestamp to each update without engaging in synchronous coordination
with other partitions, or with \thesystem. We will explain
below how this can be easily achieved. These timestamps must
satisfy two properties.

\begin{property}
\label{prop:readfrom}
If an update $u_j$ causally depends on a second update $u_i$, then
 the timestamp assigned to $u_j$ ($u_j.ts$) is strictly greater than $u_i.ts$.
\end{property}

\begin{property}
\label{prop:monotonicity}
For two updates $u_i$ and $u_j$ received by \thesystem from partition $p_n$, if $u_i$
is received before $u_j$ then $u_j.ts$ is
strictly greater than $u_i.ts$.
\end{property}

These two properties imply that updates are causally ordered across all
partitions and that once \thesystem receives an update coming from
a partition $p_n$, no update with smaller timestamp will be
ever received from $p_n$. In order to ensure these properties, clients play a
fundamental role. A client $c$ maintains a local variable,
\textit{Clock}$_c$, that stores the largest timestamp seen during its
session. This clock value captures client's causal dependencies and
it is included in every update request. As described below, partitions
compute update timestamps taking into account the value of client clocks.

The protocol assumes that each partition $p_n$ is equipped with a
physical clock. Clocks are loosely synchronized by a time
synchronization protocol such as NTP\,\cite{ntp}. The correctness of
the protocol does not depend on the clock synchronization
precision and can tolerate clock drifts. However, as discussed later,
large clock drifts could have a negative impact on the protocol
performance (in particular, on how fast the datacenter can ship
updates to remote datacenters). To avoid this limitation,
our protocol uses hybrid clocks\,\cite{HLC}, which have been shown to
overcome some of the limitations of simply using physical time.

We now describe how events are handled by clients,
partitions and \thesystem (Algs.~\ref{alg:client},~\ref{alg:server},
and~\ref{alg:system} respectively).

\begin{algorithm}[t]
  \begin{algorithmic}[1]
\footnotesize
\Function{read}{\textit{Key}}
        \State $\langle$\textit{Value}, \textit{Ts}$\rangle\gets$ \Call{kv\_get}{\textit{Key}}
        \State \Return $\langle$\textit{Value}, \textit{Ts}$\rangle$
    \EndFunction
    \vspace{1mm}
	\Function{update}{\textit{Key}, \textit{Value}, \textit{Clock}$_c$}
        \State \textit{MaxTs}$_n\gets$ \Call{max}{\textit{Clock}$_n$, \textit{Clock}$_c +1$,
          \textit{MaxTs}$_n+1$}\label{comptimestamp}
        \State \Call{kv\_put}{\textit{Key}, $\langle$\textit{Value}, \textit{MaxTs}$_n \rangle$}
        \State $u_j$ $\gets\langle$ \textit{Key}, \textit{Value}, \textit{MaxTs}$_n,p_n\rangle$
        \State {\bf send} \Call{add\_op}{$u_j$} {\bf to}
        \thesystem\label{addopserver}
        \State \Return \textit{MaxTs}$_n$
    \EndFunction
\vspace{1mm}
\Function{heartbeat}{}\Comment{Every $\Delta$ time}\label{inithb}
\If{\textit{Clock}$_n\geq $ \textit{MaxTs}$_n + \Delta$}
\State {\bf send} \Call{heartbeat}{$p_n$, \textit{Clock}$_n$} {\bf to}\label{endhb}
        \thesystem
\EndIf
    \EndFunction
  \end{algorithmic}
  \caption{Operations at partition $p_n$}
  \label{alg:server}
\end{algorithm}

\smallskip
\noindent{\bf Read} A client $c$ sends a read request on a data item (identified
by \textit{Key}) to the responsible partition $p_n$ (Alg.~\ref{alg:client},
line~\ref{clientread}). When $p_n$ receives the request, it fetches
the \textit{Value} and the timestamp \textit{Ts} that is locally stored for
\textit{Key} and returns both to the client. \textit{Ts} is the
timestamp assigned by $p_n$ to the update operation that generated
the current version. After receiving the pair
$\langle$\textit{Value}, \textit{Ts}$\rangle$, the client computes the
maximum between \textit{Clock}$_c$ and \textit{Ts} (Alg.~\ref{alg:client},
line~\ref{maxclient}) to ensure that the read operation is included in
its causal history.

\smallskip
\noindent{\bf Update} A client $c$ sends an update request operation to 
the responsible partition $p_n$ of the object being updated. Apart
from the \textit{Key} and \textit{Value}, the request includes client's clock
\textit{Clock}$_c$ (Alg.~\ref{alg:client},
line~\ref{clientupdate}). When $p_n$ receives the request, it first
computes the timestamp of the new update (Alg.~\ref{alg:server},
line~\ref{comptimestamp}). This is computed by taking the maximum
between \textit{Clock}$_n$ (physical time), the maximum timestamp ever
used by $p_n$ (\textit{MaxTs}$_n$) plus one and \textit{Clock}$_c$
(client's clock) plus one. This ensures that the timestamp is greater
than both \textit{Clock}$_c$ and any other update timestamped by
$p_n$. Then, $p_n$ stores the \textit{Value} and the recently computed
timestamp in the local key-value store and asynchronously sends the
operation to the \thesystem service. Finally, $p_n$ returns update's
timestamp to the client who updates \textit{Clock}$_c$ with it, since
it is guaranteed to be greater than its current one.

\smallskip
\noindent{\bf Timestamp Stability} When \thesystem receives an
operation from a given partition, it adds it to the set of non-stable
operations $Ops$ and updates the $p_n$ entry in the
\textit{PartitionTime} vector with operation's timestamp
(Alg.~\ref{alg:system}, lines~\ref{initadd}--\ref{endadd}).  A timestamp \textit{Ts} is \emph{stable} at \thesystem when one
is sure that no update with lower timestamp will be received from any
partition (i.e., when \thesystem is aware of all updates with
timestamp \textit{Ts} or smaller).  Periodically, \thesystem computes
the value of the maximum stable timestamp (\textit{StableTime}), which
is computed as the minimum of the \textit{PartitionTime} vector
(Alg.~\ref{alg:system},
line~\ref{computestable}). Property~\ref{prop:monotonicity} implies
that no partition will ever timestamp an update with an equal or
smaller timestamp than \textit{StableTime}. Thus, \thesystem can
confidently serialize all operations tagged with a timestamp smaller
than or equal to \textit{StableTime} (Alg.~\ref{alg:system},
line~\ref{getstable}). \thesystem can serialize them in timestamp
order, which is consistent to causality
(Property~\ref{prop:readfrom}), and then send them to other
geo-locations (Alg.~\ref{alg:system}, line~\ref{processstable}). Note
that non-causally related updates coming from different partitions may
have been timestamped with the same value. In this case, operations
are concurrent and \thesystem can process them in any
order. 

\smallskip
\noindent{\bf Heartbeats} If a partition $p_n$ does not receive an
update for a fixed period of time, it will send a heartbeat including
its current time to \thesystem (Alg.~\ref{alg:server},
lines~\ref{inithb}--\ref{endhb}). Thus, even if a partition $p_n$
receives updates at a slower pace than others, it will not slow down
the processing of other partitions updates at \thesystem. When
\thesystem receives a heartbeat from $p_n$, it simply updates its
entry in the \textit{PartitionTime} vector (Alg.~\ref{alg:system},
line~\ref{hbupdate}).

\begin{algorithm}[t]
  \begin{algorithmic}[1]
\footnotesize
\Function{add\_op}{$u_j$}
        \State \textit{Ops} $\gets$ \textit{Ops} $\cup$ $u_j$\label{initadd}
        \State $\langle$\textit{Key}, \textit{Value}, \textit{Ts},
        $p_n \rangle\gets$ $u_j$
        \State \textit{PartitionTime}$[p_n]\gets$ \textit{Ts}\label{endadd}
    \EndFunction
    \vspace{1mm}
	\Function{heartbeat}{$p_n$, \textit{Ts}}
        \State \textit{PartitionTime}$[p_n]\gets$ \textit{Ts}\label{hbupdate}
    \EndFunction
\vspace{1mm}
\Function{process\_stable}{}\Comment{Every $\theta$ time}
\State \textit{StableTime} $\gets$ \Call{min}{\textit{PartitionTime}}\label{computestable}
\State \textit{StableOps} $\gets$ \Call{find\_stable}{\textit{Ops}, \textit{StableTime}}\label{getstable}
\State \Call{process}{\textit{StableOps}}\label{processstable}
\State \textit{Ops} $\gets$ \textit{Ops} $\setminus$ \textit{StableOps}
    \EndFunction
  \end{algorithmic}
  \caption{Operations at \thesystem}
  \label{alg:system}
\end{algorithm}

\subsection{Discussion}
\noindent{\bf Hybrid Clocks} Our protocol combines logical and
physical time. Although \thesystem could simply use logical clocks and
still be correct, the rate at which clocks from different partitions
progress would depend on the rate in which partitions receive update
requests. This may cause \thesystem to process local updates in a
slower pace and thus increase remote visibility latencies, as the
stable time is set to the smallest timestamp received among all partitions. Differently, physical clocks naturally progress at similar
rates independently of the workload
characterization. This fact---previously exploited
by~\cite{gentlerain, cure}---makes stabilization procedures resilient
to skewed load distribution. Unfortunately, physical clocks do not progress exactly at the same rate, forcing
protocols to wait for clocks to catch up in some situations in order to ensure
correctness~\cite{clocksi, gentlerain, cure,
  clockrsm}. The logical part of the hybrid clock makes the protocol
resilient to clock skew by avoiding artificial delays
due to clock synchronization uncertainties~\cite{HLC}. Briefly, if a
partition $p_n$ receives an update request with
\textit{Clock}$_c>$ \textit{Clock}$_n$, instead of waiting until
\textit{Clock}$_n>$ \textit{Clock}$_c$ to ensure correctness
(Property~\ref{prop:readfrom}), the logical part of the hybrid clock
(\textit{MaxTs}$_n$) is moved forward. Then, when a partition
$p_n$ receives an update from any client, if the physical part \textit{Clock}$_n$ is
still behind the logical (\textit{MaxTs}$_n$), the update is tagged
with \textit{MaxTs}$_n+1$ in order to ensure clock monotonicity and
thus guarantee Property~\ref{prop:monotonicity}.

\smallskip
\noindent{\bf Correctness}
We provide an informal proof that our protocol satisfies the two
properties required by \thesystem (Properties~\ref{prop:readfrom}
and~\ref{prop:monotonicity}).

Property~\ref{prop:monotonicity} is trivial to prove. We need to
ensure that updates handled by a partition $p_n$ are tagged with
strictly increasing timestamps and that
heartbeats do not break the monotonicity. By Algorithm~\ref{alg:server}
line~\ref{comptimestamp}, $p_n$ ensures that consecutive updates are tagged
with increasing timestamps. On the other hand, heartbeats are only
sent when the physical clock at $p_n$ is greater or equal to the
lastest timestamp used to tag an update plus a fixed time
$\Delta$ (Alg.~\ref{alg:server}, line~\ref{inithb}). This ensures that a heartbeat message is always tagged with a larger
timestamp that all previously processed updates. Finally, an update
happening right after a heartbeat is always tagged with a larger
timestamp than the heartbeat's timestamp
since the physical clock ($Clock_n$) is used to compute update's
timestamp and this is assumed to increase monotonically (Alg.~\ref{alg:server}
line~\ref{comptimestamp}).

In order to prove Property~\ref{prop:readfrom} we need to prove that
the partial order derived from update timestamps is consistent with
causality. To respect causality, an update $u_j$ issued by client $c$
has to be tagged with a timestamp strictly greater than all its
previous updates and than any version previously read by that
client. \textit{Clock}$_c$, which is the clock maintained by the
client, aggregates the client's causal history in a single scalar. By
Algorithm~\ref{alg:server} line~\ref{comptimestamp}, we know that the
timestamp assigned to a client update is strictly greater than
\textit{Clock}$_c$. Thus, we only need to prove that
\textit{Clock}$_c$ is always equal or greater than all previously read
versions, ensured by Algorithm~\ref{alg:client} line~\ref{maxclient},
and that it is always greater or equal to the timestamp assigned to
its last update, ensured by Algorithm~\ref{alg:client}
line~\ref{tsupdate}.

\subsection{Fault-Tolerance}
\label{sec:faulttolerance}

In the description above, for simplicity, we have described the
\thesystem service as if implemented by a single non-replicated
server. Naturally, as any other service in a datacenter, \thesystem
must be made fault-tolerant. In fact, if \thesystem fails, the
site stabilization procedure stops, and thus,
local updates can no longer be propagated to other geo-locations. In
order to avoid such limitation, we now propose a fault-tolerant
version of \thesystem. Note that we disregard failures in datacenters,
as the problem of making data services fault-tolerant has been widely
studied and is orthogonal to our work.

In this new version, \thesystem is composed by a set of
\textit{Replicas}. Algorithm~\ref{alg:ftsystem} shows the behaviour of
a replica $e_f$ of the fault-tolerant \thesystem service. We assume
the initial set of \thesystem replicas is common knowledge: every
replica knows every other replica and every partition knows the full set
of replicas. Partitions send operations and heartbeats
(Alg.~\ref{alg:server}, lines~\ref{addopserver} and~\ref{endhb}
respectively) to the whole set of \thesystem replicas. The correctness
of the algorithm requires the communication between partitions and
\thesystem replicas to satisfy the
\emph{prefix-property}\,\cite{bayoulog}: an \thesystem replica $r_f$
that holds an update $u_j$ originating at $p_n$ also holds any other
update $u_i$ originating at $p_n$ such that $u_i.ts < u_j.ts$. This
property can be ensured with inexpensive protocols that offer only
\emph{at-least-once delivery}. Stronger properties, such as
inter-partition order or exactly-once delivery are not required to
enforce the prefix-property. Our implementation achieves the
prefix-property by having each partition to keep track of the latest timestamp
acknowledged by each of the \thesystem replicas in a vector denoted as
\textit{Ack}$_n$. Thus, to each \thesystem replica $e_f$, a
partition $p_n$ sends not only the lastest update but the set of
updates including all updates $u_j$ such that
$u_j.ts>$\textit{Ack}$_n[f]$. Upon receiving a new batch of updates
\textit{Batch} (Alg.~\ref{alg:ftsystem},
lines~\ref{initbatch}--\ref{endbatch}), $e_f$ process it---in
timestamp order---filtering out those updates already seen, and updating
both \textit{Ops}$_f$ and \textit{PartitionTime}$_f$ accordingly with
the timestamps of the unseen updates. After processing \textit{Batch},
$e_f$ acknowledges $p_n$ including the greatest timestamp observed
from updates originating at $p_n$ (\textit{PartitionTime}$_f[p_n]$). This algorithm is resilient to
message lost and unordered delivery. Nevertheless, it adds redundancy,
as replicas may receive the same update multiple
times. \S\ref{sec:optimizations} proposes a set of optimizations that aim
to reduce this overhead.

\begin{algorithm}[t]
  \begin{algorithmic}[1]
\footnotesize
\Function{new\_batch}{\textit{Batch}, $p_n$}\label{initbatch}
\ForAll{$u_j \in$ \textit{Batch}$,$\textit{PartitionTime}$_f[p_n]<u_j.ts$}
\State \textit{PartitionTime}$_f[p_n]$ $\gets$ $u_j.ts$
\State \textit{Ops}$_f$ $\gets$ \textit{Ops}$_f$ $\cup$ $u_j$
\EndFor
\State {\bf send} \Call{ack}{\textit{PartitionTime}$_f[p_n]$} {\bf to} $p_n$
\vspace{1mm}
\EndFunction \label{endbatch}
\Function{process\_stable}{}\Comment{Every $\theta$ time}
\If{\textit{Leader}$_f==e_f$}\label{popsinit}
\State \textit{StableTime} $\gets$ \Call{min}{\textit{PartitionTime}$_f$}
\State \textit{StableOps} $\gets$ \Call{find\_stable}{\textit{Ops}$_f$, \textit{StableTime}}
\State \Call{process}{\textit{StableOps}}\label{popsend}
\State \textit{Ops}$_f\gets$ \textit{Ops}$_f\setminus$ \textit{StableOps}
\State {\bf send} \Call{stable}{\textit{StableTime}} {\bf to}
        \textit{Replicas}$_f\setminus \{e_f\}$\label{sendstable}
\EndIf
    \EndFunction
\vspace{1mm}
\Function{stable}{$StableTime$}
\State \textit{StableOps} $\gets$ \Call{find\_stable}{\textit{Ops}$_f$, \textit{StableTime}}\label{initstable}
\State \textit{Ops}$_f\gets$ \textit{Ops}$_f\setminus$ \textit{StableOps}\label{endstable}
    \EndFunction
\vspace{1mm}
\Function{new\_leader}{$e_g$}
\State \textit{Leader}$_f\gets e_g$\label{newleader}
    \EndFunction
  \end{algorithmic}
  \caption{Operations at \thesystem replica $e_f$}
  \label{alg:ftsystem}
\end{algorithm}


In addition, to avoid unnecessary redundancy when exchanging metadata
among datacenters, a leader replica is elected to propagate this
information. The existence of a unique leader is not required for the
correctness of the algorithm; it is simply a mechanism to save network
resources. Thus, any leader election protocol designed for
asynchronous systems (such as
$\Omega$~\cite{Chandra:1996}) can be plugged into our implementation. A change in the leadership is notified
to a replica $e_f$ through the {\sc new\_leader} function
(Alg.~\ref{alg:ftsystem}, line~\ref{newleader}).  The notion of a
leader is used to optimize the service's operation as follows. When
the {\sc process\_stable} event is triggered, only the leader replica
computes the new stable time and processes stable operations
(Alg.~\ref{alg:ftsystem}, lines~\ref{popsinit}--\ref{popsend}). Then,
after operations have been processed, the leader sends the recently
computed \textit{StableTime} to the remaining replicas
(Alg.~\ref{alg:ftsystem}, line~\ref{sendstable}). When replica $e_f$
receives the new stable time, it removes the operations already known
to be stable from its pending set of operations, since it is certain
that those operations have been already processed
(Alg.~\ref{alg:ftsystem}, lines~\ref{initstable}--\ref{endstable}).


\section{Supporting Geo-replication}
\label{sec:georeplication}

In our previous protocol, we have shown how to unobtrusively timestamp local updates in a
partial order consistent with causality. In this section, we complete
the protocol with the necessary mechanisms to ensure
that remote updates---coming from other datacenters---are made
visible locally without violating causality. Our solution resembles protocols implemented by
other causally consistent geo-replicated storage
systems\,\cite{chainreaction, swiftcloud}. We assume a total of $M$
geo-locations, each of them replicating the full set of objects. Each
of these geo-locations uses the \thesystem service and thus
propagates local updates in a total order consistent to causal
consistency. We assume FIFO links between datacenters.


Apart from the \thesystem service, each datacenter is extended with a
\textit{receiver}. This component coordinates the execution of remote
updates. Thus, it receives remote updates coming from remote
\thesystem services (as a result of {\sc process\_stable}), and
forwards them to the local datacenter partitions when its causal
dependencies are satisfied. Standard replication
techniques\,\cite{chainreplication, paxos, primary, tutorialft} can be employed to make receivers robust to failures, as
otherwise they represent a single point of failure.

We now proceed to explain how the metadata is enriched and the changes we
need to apply to our previous
algorithms. Table~\ref{table:definitionsgeo} provides a summary of the
notation used in this section. 

\begin{table}[t!]
\centering
\footnotesize
\renewcommand{\arraystretch}{1}
  \begin{tabulary}{\columnwidth}{@{} R | L @{}}
    $M$ & Number of datacenters\\
\hline
    \textit{VClock}$_c$ & Client $c$ vector ($M$ entries)\\
\hline
    $p_n^m$ & Partition $n$ at datacenter $m$\\
    $r_m$ & Receiver at datacenter $m$\\
    \textit{SiteTime}$_m$ & Applied updates vector at $r_m$\\
    \textit{Queue}$_m$ & Queues of pending updates at $r_m$\\
\hline
    $u_j.vts$ & Update $u_j$ timestamp vector ($M$ entries) \\
  \end{tabulary}
\caption{Notation used in the geo-replicated protocol extension.}
\label{table:definitionsgeo}
\end{table}

Updates
are now tagged with a vector with an entry per datacenter, capturing
inter-datacenter dependencies. The client clock is consequently also
extended to a vector (\textit{VClock}$_c$).  We could easily adapt our protocols to
use a single scalar, as in\,\cite{gentlerain}. Nevertheless, vector
clocks make a more efficient tracking of causal dependencies
introducing no false dependencies across datacenters, which reduces
the update visibility latency, at the cost of slightly increasing the
storage and computation overhead. This overhead, unlike
in\,\cite{cure}, is negligible in our protocol as \thesystem
allows for trivial dependency checking procedures. Note that the lower-bound update
visibility latency for a system relying on vector clocks is the
latency between the originator of the update and the remote
datacenter, while with a single scalar it is the latency to the
farthest datacenter regardless of the originator of the update.
\begin{algorithm}[t]
  \begin{algorithmic}[1]
\footnotesize
\Function{new\_update}{$u_j$, $k$}
       \State \textit{Queue}$_m[k]\gets [$\textit{Queue}$_m[k] | u_j]$\Comment{add to tail}
\EndFunction
\vspace{1mm}
\Function{check\_pending}{}\Comment{Every $\rho$ time}
          \State $\langle$\textit{Queue}$_m$,
          \textit{SiteTime}$_m$$\rangle\gets$ \Call{flush}{$1$, \textit{Queue}$_m$, \textit{SiteTime}$_m$}\label{flush}
\EndFunction
\vspace{1mm}
\Function{flush}{$k$, \textit{Queue}$_m$, \textit{SiteTime}$_m$}
           \If{$k>M$}
                 \State \Return $\langle$\textit{Queue}$_m$, \textit{SiteTime}$_m$$\rangle$
           \ElsIf{$k=m$}\Comment{}
                 \State \Call{flush}{$k+1$,
                   \textit{Queue}$_m$, \textit{SiteTime}$_m$}
           \Else
                 \State $u_j\gets$\Call{head}{\textit{Queue}$_m[k]$}
                 \If{$\forall d\in
                   M\setminus\{m,k\},$\textit{SiteTime}$_m[d]\geq
                   u_j.vts\lbrack d\rbrack$}\label{condition2}
                       \State $p_n^m\gets$ \Call{responsible}{$u_j.key$}
                       \State {\bf send} \Call{apply}{$u_j$} {\bf to} $p_n^m$
                       \State {\bf receive} \textit{ok} {\bf from} $p_n^m$
                       \State \textit{SiteTime}$_m[k]\gets u_j.vts[k]$
                       \State \Call{pop}{\textit{Queue}$_m[k]$}
                       \State \Call{flush}{$1$, \textit{Queue}$_m$, \textit{SiteTime}$_m$}\label{flush}
           \Else
                 \State \Call{flush}{$k+1$,
                   \textit{Queue}$_m$, \textit{SiteTime}$_m$}
           \EndIf
           \EndIf
    \EndFunction
  \end{algorithmic}
  \caption{Operations at $r_m$}
  \label{alg:receiver}
\end{algorithm}

\smallskip
\smallskip
\noindent{\bf Update} When a client $c$ issues an update operation, it piggybacks its
\textit{VClock}$_c$ summarizing both local and remote dependencies. A
partition $p_n$ computes $u_j$ vector timestamp ($u_j.vts$) as
follows. First, the local entry of the vector $u_j.vts\lbrack
m\rbrack$ is computed as the maximum between \textit{Clock}$_n$,
\textit{MaxTs}$_n+1$ and \textit{VClock}$_c\lbrack m\rbrack+1$,
similarly to Algorithm~\ref{alg:server},
line~\ref{comptimestamp}. This permits \thesystem to still be able to
causally order local updates based on $u_j.vts\lbrack
m\rbrack$. Second, the remaining entries (remote datacenter entries)
are assigned to their sibling entries in \textit{VClock}$_c$. When the
operation is completed, $p_n$ returns $u_j.vts$ to the client who can
directly substitutes its \textit{VClock}$_c$ since $u_j.vts$ is known
to be strictly greater than \textit{VClock}$_c$.

\smallskip
\noindent{\bf Read} Read operations execute as in Algorithms~\ref{alg:client}
and~\ref{alg:server}. The only difference is that the returned
timestamp is a vector instead of a scalar. Thus, in order to update
\textit{VClock}$_c$, a client $c$ applies the {\sc max} operation per
entry.

\smallskip
\noindent{\bf Update Propagation} The site stabilization procedure proceeds as before, totally
ordering local updates based on the local entry of their vector
timestamp ($u.vts\lbrack m\rbrack$). \thesystem propagates local
updates to remote datacenters in $u.vts\lbrack m\rbrack$ order. Each
update piggybacks its $u.vts$. 

\smallskip
\noindent{\bf Remote Update Visibility} Algorithm~\ref{alg:receiver}
details receivers' operation.  A receiver $r_m$ maintains two
important pieces of state: a queue of
pending updates per remote datacenter (\textit{Queue}$_m[k]$), and a
vector with an entry per remote datacenter (\textit{SiteTime}$_m$) indicating the latest update operation
locally applied from each of the remote datacenters.
When $r_m$ receives a remote update
$u_j$ coming from datacenter $k$, it simply adds it to its
corresponding queue. Periodically, $r_m$ triggers the {\sc
  check\_pending} function (Algorithm~\ref{alg:receiver}
line~\ref{flush}). This function ensures, by means of the tail recursive {\sc
  flush} function, that no pending operation is left unexecuted. Two conditions have to be satisfied
before sending an update $u_j$ to local partitions:
\begin{inparaenum}[\it (i)]
\item 
  all previously received updates coming from $k$ have already been 
  applied locally; and
\item
  $u_j$ dependencies, which are subsumed in $u_j.vts$, are visible locally.
\end{inparaenum}
Both conditions are trivially checked by relying on the information
subsumed in \textit{Queue}$_m$ and
\textit{SiteTime}$_m$. When
a pending operation $u_j$ originating at $k$ is applied, both
\textit{Queue}$_m[k]$ and \textit{SiteTime}$_m[k]$ are updated consequently.

\section{Optimizations}
\label{sec:optimizations}
We propose a set of optimizations that aim at enabling \thesystem to handle even
heavier loads.

\smallskip
\noindent{\bf Communication Patterns} \thesystem constantly receives operations and heartbeats from
partitions. This is an all-to-one 
communication schema and, if the number of partitions is large, it
may not scale in practice. In order to overcome
this problem and efficiently manage a large number of partitions, two
simple techniques have been used:
\begin{inparaenum}[\it (i)]
\item 
  build a propagation tree among partition servers; and
\item
  batch operations at partitions, and propagate them to
  \thesystem only periodically. 
\end{inparaenum}
Both techniques are able to reduce the number of
messages received by \thesystem per unit of time at the cost of a slight
increase in the stabilization time.

\smallskip
\noindent{\bf Separation of Data and Metadata} In the protocols
described before, partitions send updates (including the update value)
to the \thesystem service, which is responsible for eventually
propagating them to remote datacenters. This can limit the maximum
load that \thesystem can handle and become a bottleneck due to the
potentially large amount of data that has to be handled. In order to
overcome this limitation, we decouple data from metadata.

In our prototype, for each update operation, partitions generate a
unique update identifier ($u.id$), composed of the local entry of the
update vector timestamp ($u.vts[m]$) and the object identifier
(\textit{Key}). We avoid sending the value of the update to
\thesystem. Instead, partitions only send the unique identifier $u.id$
together with the partition id ($p_n^m$). \thesystem is then only
responsible for handling and propagating these lightweight
identifiers, while the partitions itself are responsible for
propagating (with no order delivery constraints) the update values together with $u.id$ to its sibling
partitions in other datacenters. A receiver $r_m$ proceeds as before,
but a partition $p_n^m$ can only execute the remote operation once it
has received both the data and the metadata. This technique slightly
increases the computation overhead at partitions, but it allows
\thesystem to handle a significantly heavier load independently of
update values.

\section{Implementation}
\label{implementation}

The \thesystem service has been implemented in the \texttt{C++}
programming language and integrated with a version of Riak KV\,\cite{riakkv}, a weakly
consistent datastore used by many companies offering cloud-based
services including bet365 and
Rovio. Since Riak KV is implemented in Erlang, we first attempted to
build \thesystem using the
\texttt{Erlang/OTP} framework, but unfortunately we rapidly reached a bottleneck in our early
experiments due to the
inefficiency of \texttt{Erlang} data structures. Note that for
\thesystem to work, we need to store a potentially very large number
of updates, coming
from all logical partitions composing a datacenter, and
periodically traversed them in timestamp order when a new stable time
is computed. Inserting and traversing this (ordered) set of updates was
limiting the maximum load that \thesystem could
handle. The \texttt{C++} implementation does not suffer from these performance limitations.

At its core, \thesystem is implemented using a \emph{red-black tree}\,\cite{rbtree}, 
a self-balancing binary search tree optimized for insertions and deletions, which
guarantees logarithmic search, insert and delete cost, and linear
in-order traversal cost, a critical operation for \thesystem. For our particular case, the
\emph{red-black tree} turned out
to be more efficient than other self-balancing binary search trees
such as AVL trees\,\cite{avltree}. 

Furthermore, in order to fully explore the capacities of \thesystem
and experimentally demonstrate our hypothesis, we have integrated
\thesystem with a causally consistent geo-replicated datastore
implementing the protocol presented in \S\ref{sec:thesystem} and
\S\ref{sec:georeplication}. Our prototype, namely \theprototype, is
built as a variant of Riak KV\,\cite{riakkv}, and includes the
optimizations discussed in \S\ref{sec:optimizations}. Since the open source
version does not support replication across Riak KV clusters, we have
also augmented the open source version of Riak KV with geo-replication
support.

\section{Evaluation}
\label{sec:evaluation}

Our main goal with the evaluation is to show that \thesystem does not
suffer from the limitations of the competing approaches. Therefore, we
compare \thesystem both with approaches based on sequencers and based
on global stabilization. We recall that the main disadvantage of
sequencers is to throttle throughput, because they operate in the
critical path of local clients. Therefore, we aim at showing that
\thesystem does not compromise the intra-datacenter concurrency and
can reach higher throughput that sequencer-based
approaches. Conversely, the expensiveness of the global stabilization
approach forces designers to favour either throughput or remote update
visibility latencies. Thus, we also aim at showing that \thesystem
optimizes both.

\smallskip
\noindent{\bf Experimental Setup} The experimental test-bed used is a private cloud composed by a set of
virtual machines deployed over 20 physical machines (8 cores and 40 GB
of RAM) connected via a Gigabit switch. Each VM, which runs Ubuntu
14.04, and is equipped with 2 (virtual) cores, 10GB disk and 9GB
of RAM memory; is allocated in a different physical machine. Before running
each experiment, physical clocks are synchronized using the NTP
protocol\,\cite{ntp} through a near NTP server.

\smallskip
\noindent{\bf Workload Generator} Each client VM runs its own instance of a custom version of Basho
Bench\,\cite{bashobench}, a load-generator and benchmarking tool to
conduct accurate and repeatable performance tests. For each
experiment, we deploy as many client instances as possible without overloading
the system. Latencies across
datacenters are emulated using \texttt{netem}\,\cite{netem}, a Linux network
emulator tool. 

In our experiments, unless specified, we use the following
parameters. Values used in operations are a
fixed binary of 100
bytes. We use a uniform key distribution across a total of 100k keys
(objects). The ratio of reads and updates is varied depending on the
experiment. Before running the experiments, we populate the database. Each
experiment runs for more than 6 minutes. In our results, the first and the last
minute of each experiment is ignored to avoid experimental artifacts.

\subsection{\thesystem Throughput}
\label{sec:upperbound}

We  report on a number of experiments that aim at:
\begin{inparaenum}[\it (i)]
\item 
  measuring the maximum load that our efficient
  implementation of \thesystem can handle, varying the number of
  partitions connected to it; and
\item
  assessing how replication and failures affect \thesystem 's  performance.
\end{inparaenum}

For comparison, these experiments also show the maximum load that a
traditional sequencer can handle. Our implementation of a sequencer
mimics traditional implementations\,\cite{swiftcloud,
  chainreaction}. In every update operation, datacenter partitions
synchronously request a monotonically increasing number to the
sequencer before returning to the client. We have also implemented a
fault-tolerant version of the sequencer based on chain
replication\,\cite{chainreplication}: Replicas of the sequencer are
organized in a chain. Partitions send requests to the head of the
chain. Requests traverse the chain up to the tail. When the tail
receives a request, it replies back to the partition, which in turn
returns to the client.

In order to stretch as much as possible the implementation,
circumventing potential bottlenecks in the system, we
directly connect clients to \thesystem, bypassing the data
store. Thus, each client simulates a different partition in a multi-server datacenter. This allowed us to
emulate very large datacenters, with much more servers
than the ones that were at our disposal for this experiments, and
overload \thesystem in a way that would be otherwise impossible with
our testbed.


\smallskip
\noindent{\bf Throughput Upper-Bound} We first compare the non fault-tolerant version of the \thesystem
against a non fault-tolerant implementation of a sequencer. The
\thesystem implementation used for the experiment is configured to
batch updates and only send them to \thesystem after 1ms.

\begin{figure}[t]
\centering
    \includegraphics[width=\thefiguresize]{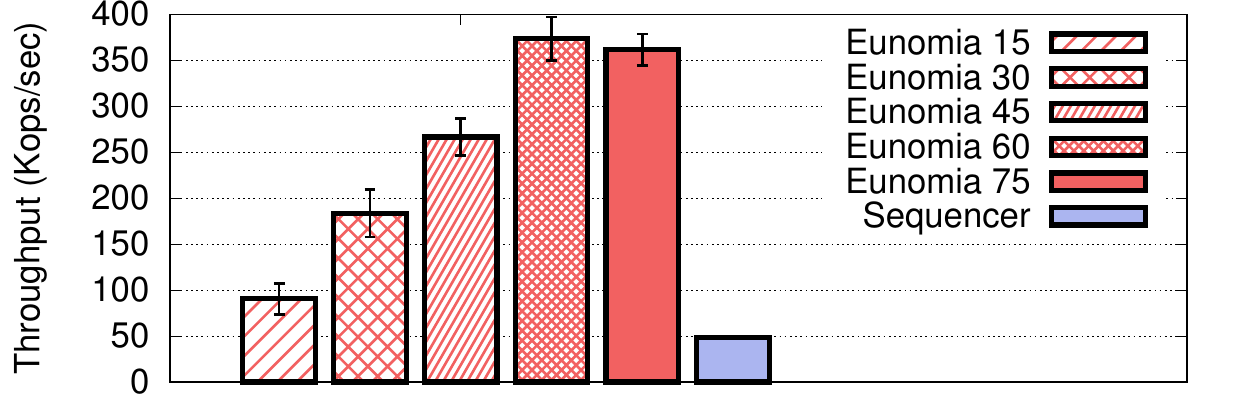}
    \caption{Maximum throughput achieved by \thesystem and an
      implementation of a sequencer. We vary the number of partitions
      that propagate operations to \thesystem.}
    \label{fig:systempartitions}
\end{figure}

Figure~\ref{fig:systempartitions} plots the maximum throughput
achieved by both services.  As results show, \thesystem maximum
throughput is reached when having 60 partitions issuing operations
eagerly (with zero waiting time between operations). We observe that
\thesystem is able to handle almost an order of magnitude more
operations per second than a sequencer (more precisely, $7.7$ times
more operations, exceeding 370kops while the sequencer is saturated at
48kops). Considering that according to our experiments, a single
machine in a Riak cluster is able to handle approximately 3kops per
second, results confirm that sequencers limit intra-datacenter
concurrency and can easily become a bottleneck for medium size
clusters (i.e, for clusters above 150 machines, the sequencer would be
the limiting factor of system performance), even assuming a read
dominant (9:1) workload, a common workload for internet-based
services. Nevertheless, under the same workload assumptions, more
than a thousand machines could be used before saturating \thesystem.

Another advantage of \thesystem in comparison to sequencers is that
batching is not in client's critical path. Thus, \thesystem's
throughput can be further stretched by increasing the batching time
(while slightly increasing the remote update visibility
latency). Such stretching cannot be easily achieved with sequencers,
as any attempt to batch requests at the sequencer blocks clients.

A final conclusion can be drawn from this experiment: \thesystem
maximum capacity does not significantly varies with the number of
partitions. Although we hit the maximum load with 60 partitions, we run
an extra experiment increasing the number to 75 to see if this
negatively impacts \thesystem's performance and we observed a very
similar throughput. The reason is that the bottleneck of our
\thesystem implementation is the propagation to other geo-locations
rather than the handling of operations. This confirms that the use of
a red-black self-balancing search tree was an
appropriate design choice.

\begin{figure}[t]
\centering
    \includegraphics[width=\thefiguresize]{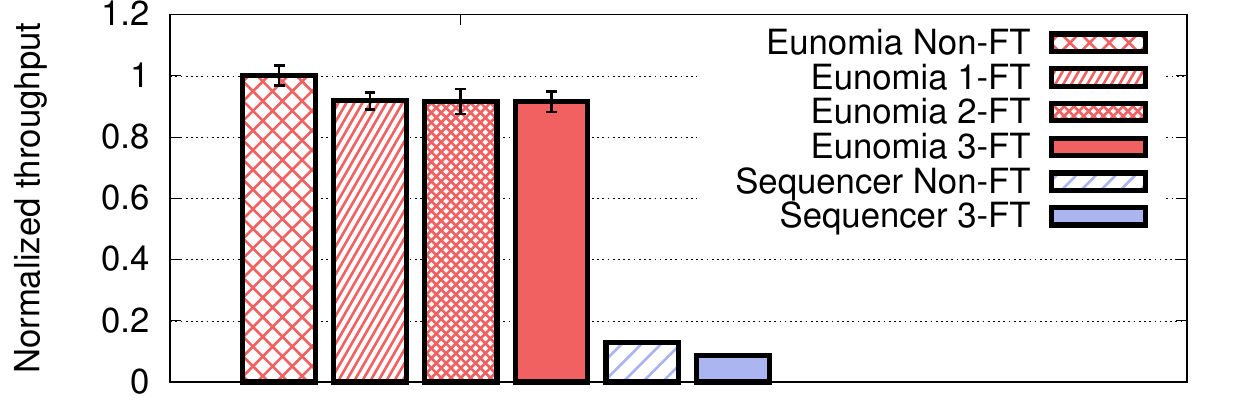}
    \caption{Maximum throughput achieved by a fault-tolerant version
      of \thesystem and sequencers. Non-FT denotes non
      fault-tolerant versions while 1-, 2-, and 3-FT denote
      fault-tolerant versions with 1, 2, and 3 replicas}
    \label{fig:systemft}
\end{figure}

\smallskip
\noindent{\bf Fault-Tolerance Overhead} In the following experiments
we measure the overhead introduced by the
fault-tolerant version of \thesystem. Figure~\ref{fig:systemft} compares the maximum throughput
achievable by \thesystem when increasing the number of replicas up to
three. For completeness, the plot also includes the throughput for a
non fault-tolerant sequencer and its fault-tolerant version with a
chain of three replicas. We normalized the throughput against the non fault-tolerant
version of \thesystem. As results show, the fault-tolerant version of
\thesystem only adds a small overhead (roughly 9\% penalty)
independently on the number of replicas. We expect this overhead to
increase as the number of replicas increases, but we consider three
replicas to be a realistic number. On the other hand, adding
fault-tolerance to the sequencer version adds a penalty of almost
33\%, thus being more expensive proportionally. The reason for this
difference is that  \thesystem replicas do not
need to coordinate as their results are independent of relative order
of inputs, while sequencer replicas need to coordinate to avoid
providing inconsistent sequence numbers.

\begin{figure}[t]
\centering
    \includegraphics[width=\thefiguresize]{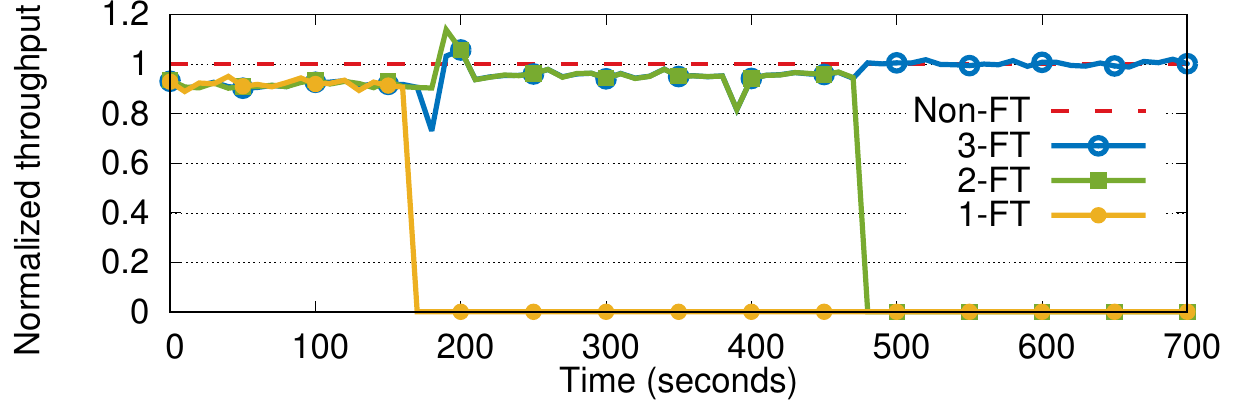}
    \caption{Impact of failures in \thesystem.}
    \label{fig:systemfaults}
\end{figure}

\smallskip
\noindent{\bf Impact of Failures} Finally, we experiment injecting failures into
\thesystem. Figure~\ref{fig:systemfaults} plots the results normalized
against the non fault-tolerant \thesystem (Non-FT line). We compare
\thesystem with one, two, and three replicas. As the figure shows, at
the beginning of the experiment, all three versions produce similar
throughput (confirming Figure~\ref{fig:systemft} results). After 160
seconds, we crash one replica. As expected, the throughput of 1-FT
drops to zero since no more replicas are available. The rest of the
versions (2-FT and 3-FT), after a short period of fluctuation,
slightly increase their throughput up to 95\% of the Non-FT version
throughput. Finally, after 210 more seconds (at 470), we crash a
second replica. Again, the 2-FT as expected drops its throughput to
zero. The 3-FT version, this time almost without fluctuations, is
capable of achieving the maximum throughput in few seconds. These
results demonstrate that failures have negligible impact in
\thesystem. Note that sometimes the multi-replica version go beyond
the Non-FT line because the Non-FT line is drawn by
computing the average.

\subsection{Experiments with Geo-Replication}
\label{sec:expstore}

We now report on a set of experiments offering evidence that a
causally consistent geo-replicated datastore built using \thesystem is
capable of providing higher throughput and better quality-of-service
than previous solutions that avoid the use of local sequencers.

For this purpose, we have implemented GentleRain\,\cite{gentlerain} and
a variation of it that uses vector clocks instead of a single scalar
to enforce causal consistency across geo-locations. The
latter resembles the causally consistency protocol implemented by
Cure\,\cite{cure}. Both approaches are sequencer-free
implementations that rely on a global stabilization procedure
in order to apply operations in remote locations consistently with
causality. For this, sibling partitions across datacenters have to
periodically send heartbeats, and each partition within a datacenter has
to periodically compute its local-datacenter
stable time. In our experiments, we set the time interval of this events
to 10 ms and 5 ms respectively unless otherwise specified. These values are in
consonance to the ones used by the authors of these works. Both approaches
are implemented using the codebase of \theprototype and thus
integrated with Riak KV.

In most of our experiments, we deploy 3 datacenters, each of them
composed of 8 logical partitions balanced across 3 servers. The
emulated round-trip-times across datacenters are 80ms
between $dc_1$ and
both $dc_2$ and $dc_3$, and 160ms between $dc_2$ and $dc_3$. These latencies
are approximately the round-trip-times between Virginia, Oregon and
Ireland regions of Amazon EC2.

\subsubsection{Throughput}
In the following experiments, we measure the throughput provided by
\theprototype, GentleRain, Cure, and an eventually consistent
multi-cluster version of Riak KV. Note that the latter does not
enforce causality, and thus partitions execute remote updates as soon
as they are received. Therefore, the comparison of \theprototype with Riak
KV allows to assess the overhead induced by \thesystem for providing
causal consistency. As discussed below, this overhead is very small.

\begin{figure}[t]
\centering
    \includegraphics[width=\thefiguresize]{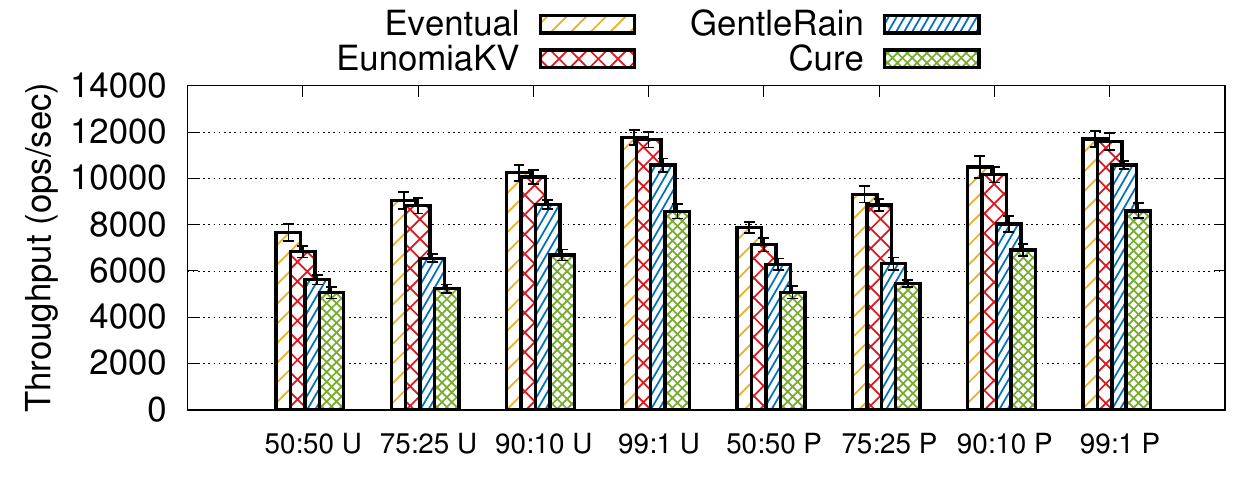}
    \caption{Throughput comparison between \theprototype and
      state-of-the-art sequencer-free solutions.}
    \label{fig:kvthput}
\end{figure}

We experiment with both uniform and power-law
key distributions, denoted with U and P respectively in Figure~\ref{fig:kvthput}. For each of them, we vary the read:write
ratio (99:1, 90:10, 75:25 and 50:50). These ratios are representative of real
large internet-based services workloads.
As shown by Figure~\ref{fig:kvthput}, the
throughput of all solutions decreases as we increase the percentage of
updates. Nevertheless, \theprototype always
provides a comparable throughput to eventual consistency. Precisely,
on average, \theprototype only drops 4.7\% of throughput,
being extremely close in read intensive workloads (1\%
drop). Differently, GentleRain and Cure are always significantly below both eventual consistency
(and \theprototype). This is due to the cost of the global
  stabilization procedure. Note that the throughput
difference between GentleRain and Cure is caused by the overhead
introduced by the metadata enrichment procedure of the latter (as
discussed in \S\ref{sec:georeplication}). Based on our experiments, it
is possible to conclude that the absolute number of updates per unit
of time is the factor that has the largest impact in \theprototype
(rather than key contention).

\subsubsection{Remote Update Visibility}

To compare the quality-of-service that can be provided by
\theprototype, GentleRain, and Cure, we measure remote update
visibility latencies. In \theprototype, we measure the time interval
between the data arrival and the instant in which the update is
executed at the responsible partition. Note that, for an update to be
applied, a datacenter needs to have access to the metadata (in our
case, provided by \thesystem) and check that all of its causal
dependencies have also been previously applied locally. In our
implementation, partitions ship updates immediately to remote
datacenters. Therefore, we have observed that updates are always
locally available to be applied by the time metadata indicates that
its causal dependencies are already satisfied locally. Although other
strategies could be used to ship the
payload of the updates, this has a crucial advantage for the
evaluation of \thesystem: under this deployment the update visibility
latency is exclusively influenced by the performance of the metadata
management strategy, including the stabilization delay incurred at the
originating datacenter.

On the other hand, for GentleRain and Cure, we measure
the time interval between the arrival of the remote operation to the
partition and when the global stabilization procedure
allows its visibility. Note that all values presented in the figures
already factor-out the network latencies among datacenters
(which are the same for all protocols); thus numbers capture only the
artificial artifacts inherent to the different approaches.

\begin{figure}[t]
\centering
    \includegraphics[width=\thefiguresize]{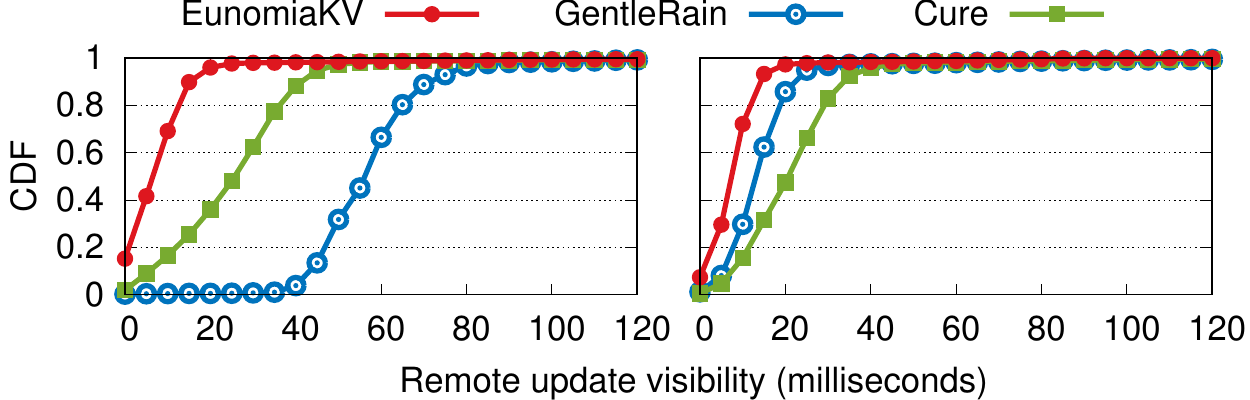}
    \caption{Left: from $dc_1$ to $dc_2$ (40ms trip-time). Right: from $dc_2$
      to $dc_3$ (80ms trip-time).}
    \label{fig:updatevisibility}
\end{figure}

Figure~\ref{fig:updatevisibility} (left plot) shows the cumulative distribution of
the latency before updates originating at $dc_1$ become visible at $dc_2$. We
observe that \theprototype offers, by far, the best remote update
visibility latency. In fact, for almost 95\% of remote updates,
\theprototype only adds 15ms extra delay. On the other hand, with
GentleRain and Cure the extra delay goes up to 80ms and 45ms
respectively for the same amount of updates. Unsurprisingly,
GentleRain extra delay is larger than Cure's because of the amount of
false dependencies added when aggregating causal dependencies into a
single scalar. In fact, GentleRain is not capable of making updates
visible without adding 40ms of extra delay. Again, the scalar is the
cause of this phenomenon since the minimum delay will not depend on the
originator of the update but on the travel time to the furthest
datacenter. This confirms the rationale presented in the discussion of \S\ref{sec:georeplication}.

Although both Cure and \theprototype rely on
vector clocks for tracking causal dependencies, \theprototype is able
to offer better remote update latencies because partitions are less
overloaded since checking dependencies in \theprototype is trivial due
to \thesystem. Note that in \theprototype, even 20\% of remote updates
are made visible without any extra delay, and thus reaching the
optimal remote update visibility latency.

Finally, in order to isolate the impact of GentleRain's global
  stabilization procedure independently of the metadata
size, we measure the remote update visibility latency at $dc_3$ for
updates originating at $dc_2$. As one can observe in
Figure~\ref{fig:updatevisibility} (right plot), GentleRain exhibits better
remote update latencies than Cure but still worse than
\theprototype. In this setting, vector clocks does not help reducing
latencies. Thus, the gap between Cure and GentleRain is exclusively
due to the storage and
computational overhead caused by vector clocks. Furthermore, the fact
that \theprototype still provides better latencies is, once again,
an empirical evidence that global
  stabilization procedures are expensive in practice.

\subsubsection{Impact of Stragglers}
Finally, we measure the sensitivity of
\thesystem to stragglers. For this, we conduct an experiment that aims
at assessing how much a straggle delays the visibility of updates
originating in healthy partitions.  The experiment uses three
datacenters (same setup of previous experiments) that run under
optimal conditions during 1 minute. Then, during the second minute, we
introduce a straggler. This is a partition of $dc_3$ that communicates
abnormally with its local \thesystem service. Thus, instead of
communicating every millisecond (as every other partition), it contacts
\thesystem less frequently. We have experimented with three
straggling intervals: 10, 100 and 1000ms. After the straggling period, the partition gets
healed and we run the experiment for one more
minute under optional conditions. Note that this is the worst case for
\thesystem, as if the communication problem were established between
partitions and remote datacenters, all solutions would suffer from
stragglers similarly.

\begin{figure}[t]
\centering
    \includegraphics[width=\thefiguresize]{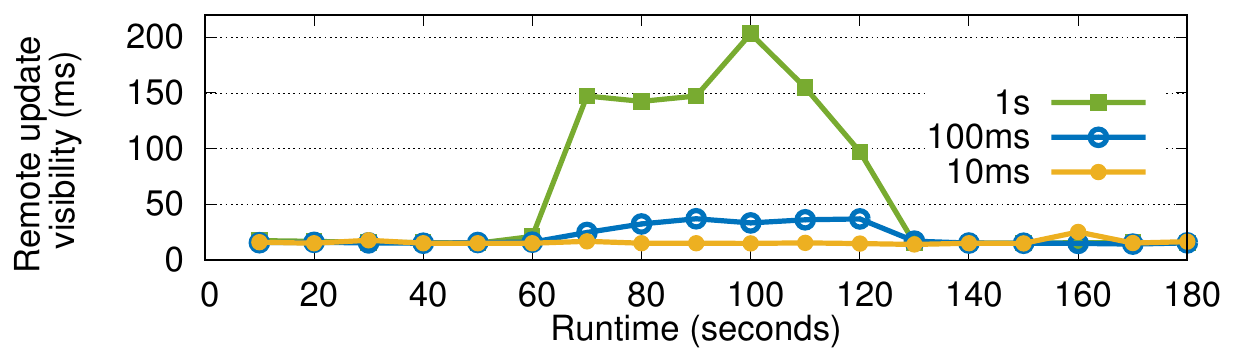}
    \caption{Stragglers impact on \thesystem.}
    \label{fig:straggler}
\end{figure}

We recall that \thesystem considers stable---thus
ready to be propagated to other datacenters---updates tagged with a
timestamp equal or smaller to the minimum among the latest timestamp
received from all local partitions. Therefore, we expect updates
originating in healthy partitions to be affected by the
communication interval set for the straggler. Figure~\ref{fig:straggler} shows remote update visibility latencies
for updates originating at $dc_3$ measured at $dc_2$. As
expected, \thesystem delays the visibility of
remote updates originating in the datacenter with the straggler
partition proportionally to the communication interval set for the
straggler.

For comparison, we also run
the same set of experiments for a sequencer-based, causally
consistent, geo-replicated storage system.\footnote{We skip the
comparison to GentleRain and Cure. We expect this systems to be
similarly affected by stragglers than \thesystem, as
they also compute the minimum among all local partitions (including
straggler) to make remote updates visible. In fact, in GentleRain, it will not only affect
updates originating at the straggler's datacenter, but also updates
originating in any other datacenter. This is caused by GentleRain's
aggressive metadata compression strategy.} In this case, the order
in which updates are propagated to remote datacenters is established
synchronously as clients interact with partitions (independently of
other partitions’ operation). Results show that, in a sequencer-based
implementation, the visibility latency of updates originating in healthy
partitions remains unaffected by stragglers. Nevertheless, as sequencers operate in
client's critical path, clients operating with stragglers
observe an increase in latency equivalent to the straggling
interval. This is unfortunate as an increase in
user-perceived latency may translate into concrete revenue loss~\cite{schurman}.

\section{Related work}
\label{sec:relatedwork}

The support for causal consistency can already be found in early
pioneer works in distributed systems, such as Bayou\,\cite{bayoulog, session}, Lazy
Replication\,\cite{lazyreplication}, and the ISIS\,\cite{birman} toolkit. Bayou offers session
guarantees and enforces causal order of write operations to an
eventually consistency data store composed by single-machine
replicas in full replication settings. Lazy Replication also ensures
that all operations are applied to replicas in causal order by having
clients maintaining a vector clock that captures their causal
past. ISIS offers a causal multicast primitive, namely \emph{cbcast}, which
provides causally ordered message delivery for group communication.

Recently, and tackling scalability challenges close to ours, multiple
weakly consistent geo-replicated data stores implementing causal
consistency across geo-locations have been proposed. We
group them into two categories: 
\begin{inparaenum}[\it (i)]
\item 
  sequencer-based solutions\,\cite{chainreaction, swiftcloud, practi};
\item
  and sequencer-free solutions\,\cite{cops, orbe, eiger, gentlerain,
  cure}.
\end{inparaenum}

\smallskip
\noindent{\bf Sequencer-based} These solutions rely on a single sequencer per datacenter to
enforce causal consistency. The sequencer is in charge of totally
ordering local updates, in a causally consistent manner, and propagate
them to remote locations. This design centralizes, thus significantly simplifying,
the arduous labor of checking remote update
dependencies. Nevertheless, the use of synchronous sequencers significantly limits
the intra-datacenter concurrency, as demonstrated by our experiments. SwiftCloud\,\cite{swiftcloud} and
ChainReaction\,\cite{chainreaction} rely on a vector
clock with an entry per datacenter to track the causal
dependencies, similarly to \theprototype. Practi\,\cite{practi}, on the
contrary, uses a
single scalar and a sophisticated mechanism of invalidations. Similar
to \theprototype, Practi separates the propagation of data and metadata. This
and the concept of \emph{imprecise} invalidations optimize Practi for
partial replication, a setting that has not yet been explored
in this work. We have shown that sequencers may get easily saturated for
medium-size clusters, while \thesystem is able to handle much heavier loads
(up to 7.7 times more).

\smallskip
\noindent{\bf Sequencer-free} There have been two major trends in this category:
\begin{inparaenum}[\it (i)]
\item 
  solutions that rely on explicit dependency check
  messages\,\cite{cops, orbe, eiger}; and
\item
  solutions based on global stabilization 
  procedures\,\cite{gentlerain, cure}.
\end{inparaenum}

COPS\,\cite{cops} and Eiger\,\cite{eiger} finely track dependencies
for each individual data item allowing full concurrency within a
datacenter. Remote updates are tagged with a list of
dependencies. When a datacenter receives a remote update, it needs to
explicitly check each dependency. This process was found to be
expensive and to limit system’s performance\,\cite{gentlerain} due to
the large amount of metadata generated. Orbe\,\cite{orbe} only
partially solves this problem by aggregating dependencies belonging to
the same logical partition into one scalar.

Alternatives that use less metadata rely on a background global
stabilization procedure\,\cite{gentlerain, cure}. This procedure
equips partitions with sufficient information to safely execute remote
updates consistently with causality. Thus, these solutions manage to
aggregate the metadata as sequencer-based solutions without relying on
an actual sequencer. As our extensive evaluation has empirically
demonstrated, global stabilization procedures are expensive in
practice. \theprototype exhibits significantly better throughput than
these solutions (therefore much better than solutions based on
explicit dependency check messages\,\cite{gentlerain}). In addition,
our evaluation have shown that \theprototype generates substantially
smaller remote update visibility latencies than GentleRain and Cure,
the two most performant solutions of the state-of-the-art.
\section{Conclusions}
\label{sec:conclusion}
We have presented a novel approach for building 
geo-replicated data stores that require updates to be
causally ordered. Our solution relies on a new service, namely
\thesystem, that abstracts the internal complexity of
datacenters, a key feature to reduce the cost of implementing causal consistency. Furthermore, unlike sequencers, \thesystem does
not limit the intra-datacenter
concurrency by performing an unobtrusive ordering of updates.

Our experimental results demonstrate that \thesystem is
able to handle very heavy loads without becoming a
performance bottleneck (up to 7.7 times more operations per second
than sequencers). Furthermore, experiments show that \theprototype (a highly
performant causally consistent geo-replicated protocol that integrates
\thesystem) provides appreciably higher throughput than GentleRain and
Cure, the two most performant solutions of the state-of-the-art. In
fact, \theprototype only adds a slight overhead (4.7\%
on average) when compared to an eventually consistent data store that
makes no attempt to enforce causality. Finally, unlike GentleRain and
Cure, \theprototype does not significantly damage
the quality-of-service provided to clients, adding
exceptionally small artificial delays on remote update visibility.

{\footnotesize \bibliographystyle{acm}
\bibliography{atc17}}

\begin{thebibliography}{10}

\bibitem{bashobench}
Basho {B}ench.
\newblock \\\url{http://github.com/basho/basho_bench}.

\bibitem{netem}
netem.
\newblock
  \\\url{http://www.linuxfoundation.org/collaborate/workgroups/networking/netem}.

\bibitem{ntp}
The network time protocol.
\newblock \\\url{http://www.ntp.org}.

\bibitem{riakkv}
Riak {KV}.
\newblock \\\url{https://github.com/basho/riak_kv}.

\bibitem{avltree}
{\sc AdelsonVelskii, M., and Landis, E.~M.}
\newblock An algorithm for the organization of information.
\newblock Tech. rep., DTIC Document, 1963.

\bibitem{causalmemory}
{\sc Ahamad, M., Neiger, G., Burns, J.~E., Kohli, P., and Hutto, P.~W.}
\newblock Causal memory: definitions, implementation, and programming.
\newblock {\em Distributed Computing 9}, 1 (1995), 37--49.

\bibitem{cure}
{\sc Akkoorath, D.~D., Tomsic, A., Bravo, M., Li, Z., Crain, T., Bieniusa, A.,
  Pregui{\c c}a, N., and Shapiro, M.}
\newblock {C}ure: Strong semantics meets high availability and low latency.
\newblock ICDCS '16.

\bibitem{chainreaction}
{\sc Almeida, S., Leit\~{a}o, J.~a., and Rodrigues, L.}
\newblock Chainreaction: A causal+ consistent datastore based on chain
  replication.
\newblock EuroSys '13.

\bibitem{primary}
{\sc Alsberg, P.~A., and Day, J.~D.}
\newblock A principle for resilient sharing of distributed resources.
\newblock ICSE '76.

\bibitem{limitations}
{\sc Attiya, H., Ellen, F., and Morrison, A.}
\newblock Limitations of highly-available eventually-consistent data stores.
\newblock PODC '15.

\bibitem{hat}
{\sc Bailis, P., Davidson, A., Fekete, A., Ghodsi, A., Hellerstein, J.~M., and
  Stoica, I.}
\newblock Highly available transactions: Virtues and limitations.
\newblock {\em Proc. VLDB Endow. 7}, 3 (Nov. 2013), 181--192.

\bibitem{potential}
{\sc Bailis, P., Fekete, A., Ghodsi, A., Hellerstein, J.~M., and Stoica, I.}
\newblock The potential dangers of causal consistency and an explicit solution.
\newblock SoCC '12.

\bibitem{indigo}
{\sc Balegas, V., Duarte, S., Ferreira, C., Rodrigues, R., Pregui\c{c}a, N.,
  Najafzadeh, M., and Shapiro, M.}
\newblock Putting consistency back into eventual consistency.
\newblock EuroSys '15.

\bibitem{birman}
{\sc Birman, K., Schiper, A., and Stephenson, P.}
\newblock Lightweight causal and atomic group multicast.
\newblock {\em ACM Trans. Comput. Syst. 9}, 3 (Aug. 1991).

\bibitem{ourpaper}
{\sc Bravo, M., Diegues, N., Zeng, J., Romano, P., and Rodrigues, L.}
\newblock On the use of clocks to enforce consistency in the cloud.
\newblock {\em IEEE Data Eng. Bull 38}, 1 (2015), 18--31.

\bibitem{Chandra:1996}
{\sc Chandra, T., Hadzilacos, V., and Toueg, S.}
\newblock The weakest failure detector for solving consensus.
\newblock {\em J. ACM 43}, 4 (July 1996), 685--722.

\bibitem{practi}
{\sc Dahlin, M., Gao, L., Nayate, A., Venkataramana, A., Yalagandula, P., and
  Zheng, J.}
\newblock Practi replication.
\newblock NSDI '06.

\bibitem{orbe}
{\sc Du, J., Elnikety, S., Roy, A., and Zwaenepoel, W.}
\newblock Orbe: Scalable causal consistency using dependency matrices and
  physical clocks.
\newblock SoCC '13.

\bibitem{clocksi}
{\sc Du, J., Elnikety, S., and Zwaenepoel, W.}
\newblock Clock-si: Snapshot isolation for partitioned data stores using
  loosely synchronized clocks.
\newblock SRDS '13.

\bibitem{gentlerain}
{\sc Du, J., Iorgulescu, C., Roy, A., and Zwaenepoel, W.}
\newblock Gentlerain: Cheap and scalable causal consistency with physical
  clocks.
\newblock SoCC '14.

\bibitem{clockrsm}
{\sc Du, J., Sciascia, D., Elnikety, S., Zwaenepoel, W., and Pedone, F.}
\newblock Clock-{RSM}: Low-latency inter-datacenter state machine replication
  using loosely synchronized physical clocks.
\newblock DSN '14.

\bibitem{rachid}
{\sc Guerraoui, R., Pavlovic, M., and Seredinschi, D.-A.}
\newblock Trade-offs in replicated systems.
\newblock {\em Data Engineering\/} (2016), 14.

\bibitem{rbtree}
{\sc Guibas, L.~J., and Sedgewick, R.}
\newblock A dichromatic framework for balanced trees.
\newblock {\em 2013 IEEE 54th Annual Symposium on Foundations of Computer
  Science 0\/} (1978), 8--21.

\bibitem{HLC}
{\sc Kulkarni, S.~S., Demirbas, M., Madappa, D., Avva, B., and Leone, M.}
\newblock Logical physical clocks.
\newblock OPODIS '14.

\bibitem{lazyreplication}
{\sc Ladin, R., Liskov, B., Shrira, L., and Ghemawat, S.}
\newblock Providing high availability using lazy replication.
\newblock {\em ACM Trans. Comput. Syst.\/} (1992).

\bibitem{paxos}
{\sc Lamport, L.}
\newblock The part-time parliament.
\newblock {\em ACM Trans. Comput. Syst. 16}, 2 (May 1998), 133--169.

\bibitem{redblue}
{\sc Li, C., Porto, D., Clement, A., Gehrke, J., Pregui{\c c}a, N., and
  Rodrigues, R.}
\newblock Making geo-replicated systems fast as possible, consistent when
  necessary.
\newblock OSDI '12.

\bibitem{cops}
{\sc Lloyd, W., Freedman, M.~J., Kaminsky, M., and Andersen, D.~G.}
\newblock Don't settle for eventual: Scalable causal consistency for wide-area
  storage with cops.
\newblock SOSP '11.

\bibitem{eiger}
{\sc Lloyd, W., Freedman, M.~J., Kaminsky, M., and Andersen, D.~G.}
\newblock Stronger semantics for low-latency geo-replicated storage.
\newblock NSDI '13.

\bibitem{cac}
{\sc Mahajan, P., Alvisi, L., and Dahlin, M.}
\newblock Consistency, availability, and convergence.
\newblock Tech. rep., University of Texas at Austin, 2011.

\bibitem{bayoulog}
{\sc Petersen, K., Spreitzer, M.~J., Terry, D.~B., Theimer, M.~M., and Demers,
  A.~J.}
\newblock Flexible update propagation for weakly consistent replication.
\newblock SOSP '97.

\bibitem{tutorialft}
{\sc Schneider, F.~B.}
\newblock Implementing fault-tolerant services using the state machine
  approach: A tutorial.
\newblock {\em ACM Comput. Surv. 22}, 4 (Dec. 1990), 299--319.

\bibitem{schurman}
{\sc Schurman, E., and Brutlag, J.}
\newblock The user and business impact of server delays, additional bytes, and
  http chunking in web search.
\newblock In {\em Velocity Web Performance and Operations Conference\/} (2009).

\bibitem{psi}
{\sc Sovran, Y., Power, R., Aguilera, M.~K., and Li, J.}
\newblock Transactional storage for geo-replicated systems.
\newblock SOSP '11.

\bibitem{session}
{\sc Terry, D.~B., Demers, A.~J., Petersen, K., Spreitzer, M.~J., Theimer,
  M.~M., and Welch, B.~B.}
\newblock Session guarantees for weakly consistent replicated data.
\newblock PDIS '94.

\bibitem{chainreplication}
{\sc Van~Renesse, R., and Schneider, F.~B.}
\newblock Chain replication for supporting high throughput and availability.
\newblock OSDI '04.

\bibitem{swiftcloud}
{\sc Zawirski, M., Pregui\c{c}a, N., Duarte, S., Bieniusa, A., Balegas, V., and
  Shapiro, M.}
\newblock Write fast, read in the past: Causal consistency for client-side
  applications.
\newblock Middleware '15.

\end{thebibliography}

\end{document}